\documentclass{article}

\usepackage[preprint,nonatbib]{neurips_2020}

\usepackage[utf8]{inputenc} 
\usepackage[T1]{fontenc}    
\usepackage{hyperref}       
\usepackage{cleveref}
\usepackage{url}            
\usepackage{booktabs}       
\usepackage{amsfonts}       
\usepackage{nicefrac}       
\usepackage{microtype}      

\usepackage{graphicx}
\usepackage{xcolor}

\usepackage{epigraph}
\title{Diptychs of human and machine perceptions}

\author{Vivien Cabannes\thanks{Equal contribution.}\\
INRIA \& ENS \& PSL\\
\href{http://www.human-machine.art/}{Human-Machine Art}
\And Thomas Kerdreux\footnotemark[1]\\
Zuse Institute, Berlin\\
\href{http://www.human-machine.art/}{Human-Machine Art}
\And Louis Thiry\\
ENS \& PSL\\
\href{http://www.human-machine.art/}{Human-Machine Art}
}

\begin{document}

\maketitle

\begin{abstract}
We propose visual creations that put differences in algorithms and humans \emph{perceptions} into perspective.
We exploit saliency maps of neural networks and visual focus of humans to create diptychs that are reinterpretations of an original image according to both machine and human attentions. 
Using those diptychs as a qualitative evaluation of perception, we discuss some crucial issues of current \textit{task-oriented} artificial intelligence.
\end{abstract}

\paragraph{Perception and creation.}
Our senses constantly receive an overwhelming amount of sensory stimuli. 
Yet, only a small amount is perceived and processed by the brain.
It results from attention mechanisms that allow us to effortlessly focus on sensory inputs that are the most likely to be useful to apprehend and operate in our environments.
As our first medium to experience the world, 
\emph{perception} has been widely studied in philosophy \cite{huxley1952doors} and psychology \cite{desimone1995neural,egeth1997visual} 
with the identification of many processes such as bottom-up \cite{treisman1980feature,yantis1984abrupt,bacon1994overriding,yantis1994stimulus} 
or top-down attention dynamics \cite{posner1980orienting,rock1981effect,duncan1984selective,tipper1994object}.
For vision specifically, eyes-tracking devices have provided rich sets of data to deepen those studies \cite{mishra2009active,ramanathan2010eye,judd2011fixations,xu2014predicting,jiang2014saliency,shen2014webpage,papadopoulos2014training}, with application in visual communication \cite{decarlo2002stylization}, art interpretation \cite{Schaeffer2019} and creation \cite{Paysant2014}.

On their end, computer scientists have relied on heuristics and concepts stemming from psychology to develop and understand image recognition algorithms.
Among others, they borrow the notion of \emph{saliency} maps to shed some light on recent neural network classifiers decisions \cite{simonyan2013deep,bach2015pixel,ribeiro2016should,selvaraju2017grad}.
Salient regions were defined as regions on which modifications are the most likely to change the output of the network.
In contrast, saliency in humans is approached via eye motions and fixation points, salient regions being those on which the human eye is the most likely to focus \cite{koch1987shifts}.

Nowadays, some creative processes integrate neural networks \cite{Gatys2015,guzdial2018co}, and artist narratives' sometimes acknowledge the algorithms as collaborators or more than mere tools \cite{White2018,cabannes2019dialog,Saeed19,Tendulkar20,Parikh20}.
Interestingly then, while perception is believed to highly influence creation\footnote{Some scholars have even inferred perception specificities of an artist based on his artwork, such as the halo effect in Van Gogh paintings, which let some to believe he was suffering from lead poisoning \cite{Ross2006}.},
neural network perception is seldom
contrasted with that of human in creative neural considerations.
Here, we propose visual creations that stir up contrasts between two modes of perception: the task-oriented artificial intelligence versus the contemplative human mind.

\vspace{-2mm}

\paragraph{Process specification.}
Our procedure consists of picture tessellation based on saliency maps.\footnote{The entire code is available online at \url{https://github.com/human-aimachine-art/perception}.} 
Given an original image, we collect saliency maps from neural networks and humans.
For a neural network, those maps correspond to the amplitude of the gradient with respect to the input of the network's output.
We compute those maps for a human according to gaze focus for which we designed an eye tracker based on facial recognition~\cite{Kazemi2014}.
Once saliency maps are computed, we sample a given number of points as if those maps were a density of probability. 
Those points are then cast into Voronoi diagrams to create a tessellation. Finally, each tile is colored according to the RGB mean of its pixels.
Some technical and aesthetic choices are discussed in Appendix \ref{app:choices}.

\begin{figure}[ht]
    \centering
    \includegraphics[width=.32\textwidth]{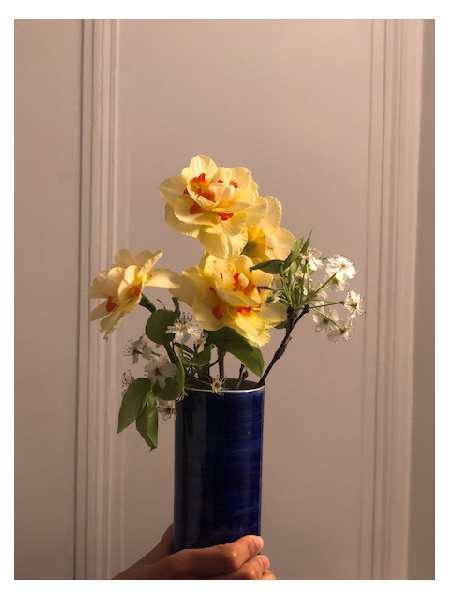}
    \includegraphics[width=.32\textwidth]{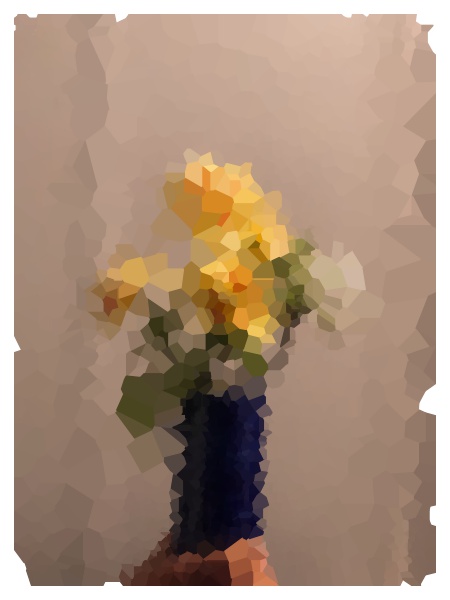}
    \includegraphics[width=.32\textwidth]{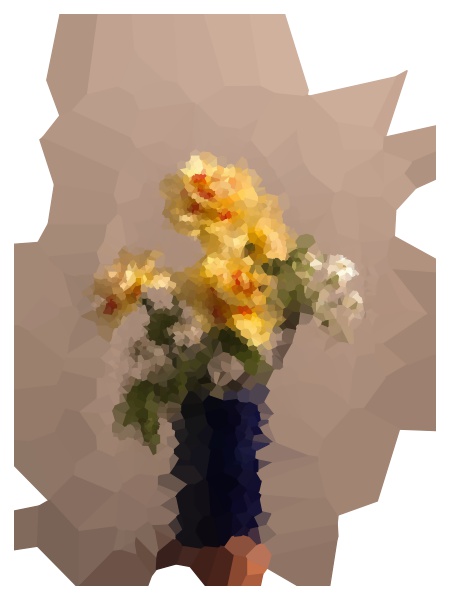}
    \includegraphics[width=.32\textwidth]{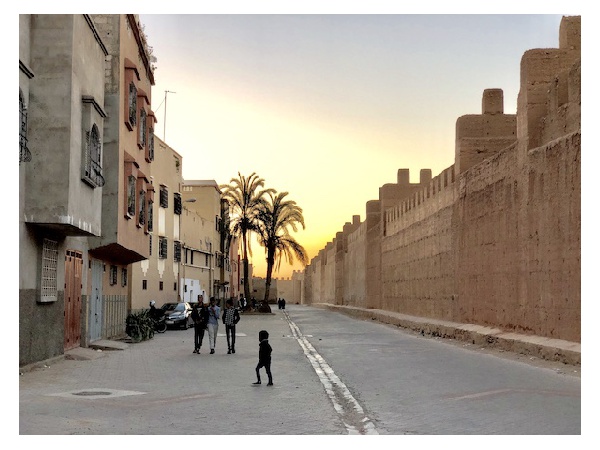} 
    \includegraphics[width=.32\textwidth]{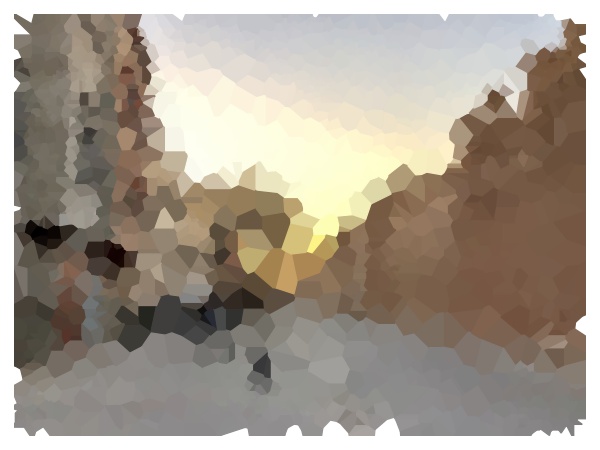} 
    \includegraphics[width=.32\textwidth]{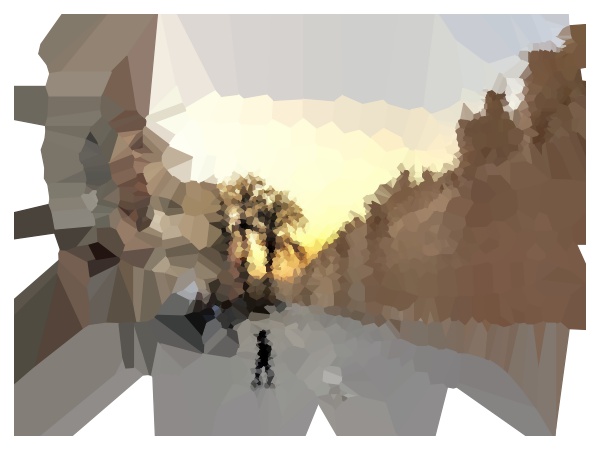}
    \caption{Diptychs of Human and Machine Perceptions. 
    From an original image (left), tessellation is achieved according to neural network attention (middle) or to human focus (right). Examples of an image with highly unambiguous content (top) and one with many focusing possibilities (bottom). 
    }
    \label{fig:triptychs}
\end{figure}

\paragraph{Diptychs discussion.}
Many notions of creativity have been developed to characterize artifacts not \emph{directly} produced by humans, could it be nature or neural networks \cite{Wyse2019}. 
However, these conceptions sometimes differ from intuitions on human creativity \cite{Colton2008}, which emphasizes the role of perception \cite{friedman2003attentional}.
By leveraging quantitative definitions of AI's attention, we hope our diptychs to provide intuitive insights on the creative agency differences between human and machine.
For instance, on~\cref{fig:triptychs}, the top example shows flowers that the machine (in the middle) recognized through the silhouette, according little attention to the details and colors of the petals, which contrasts with human focus (on the right). 
On the bottom diptychs, the human gaze eludes the buildings on the left, drawn by convergence lines to the vanishing point where lies palm trees. 
Moreover, human attention is drawn to the child playing in the foreground.
In this image, the neural network, trained for classification on \emph{ImageNet} \cite{imagenet}, misses those details that support the picture composition.
We provide more diptychs and comments in~\cref{app:additional_images}.

\begin{figure}[hb]
    \centering
    \includegraphics[width=.3\textwidth]{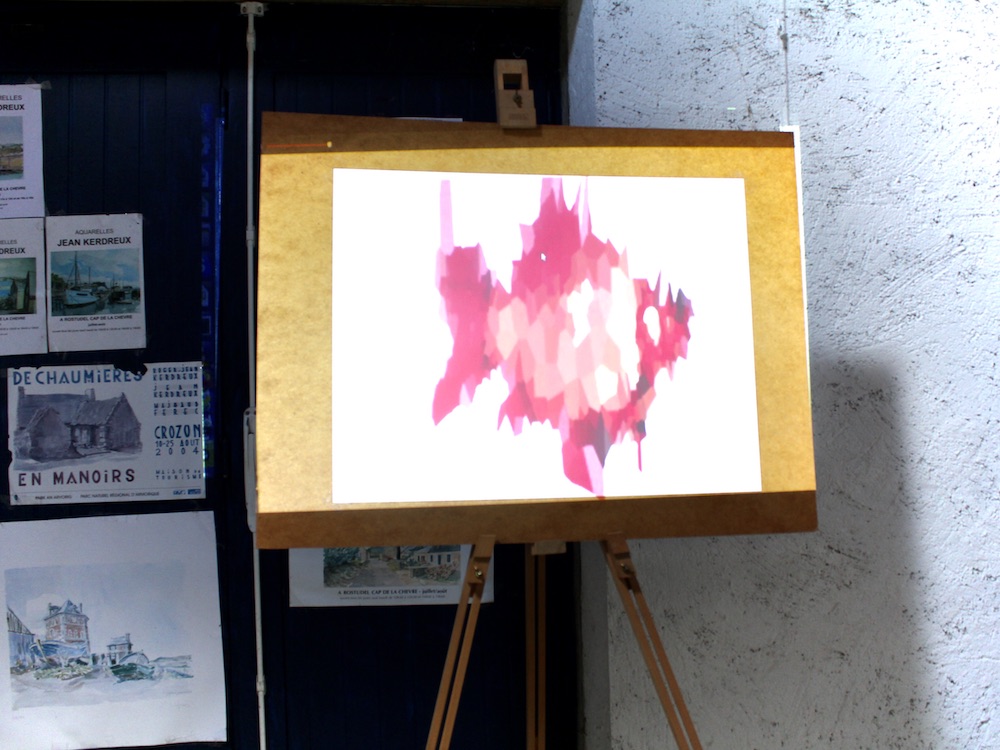}
    \includegraphics[width=.3\textwidth]{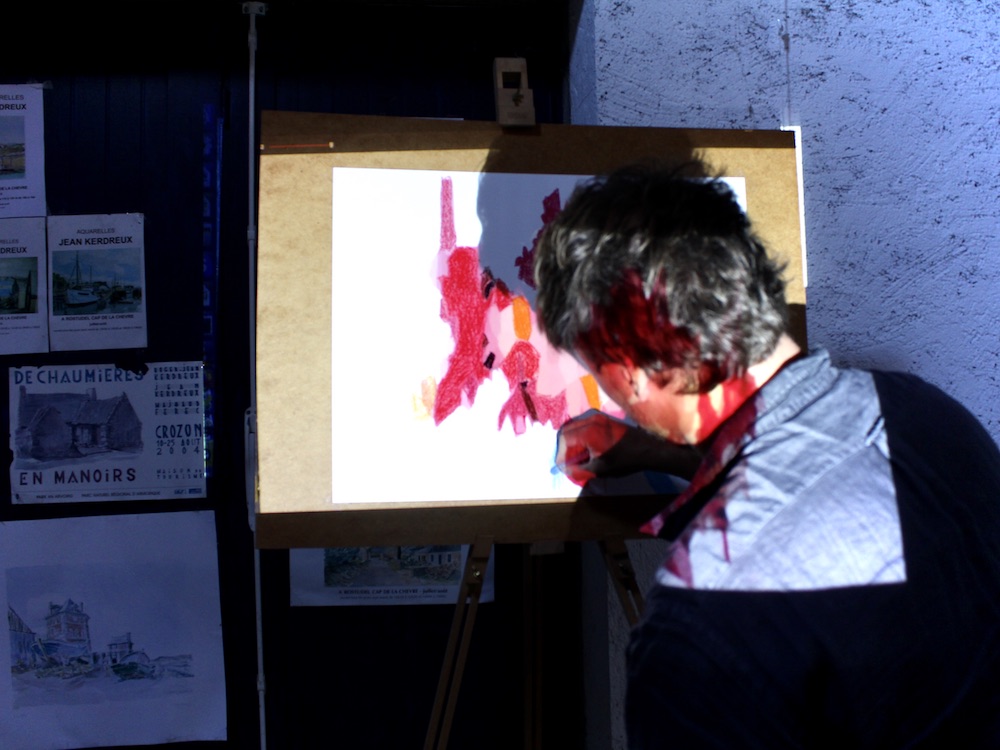}
    \includegraphics[width=.3\textwidth]{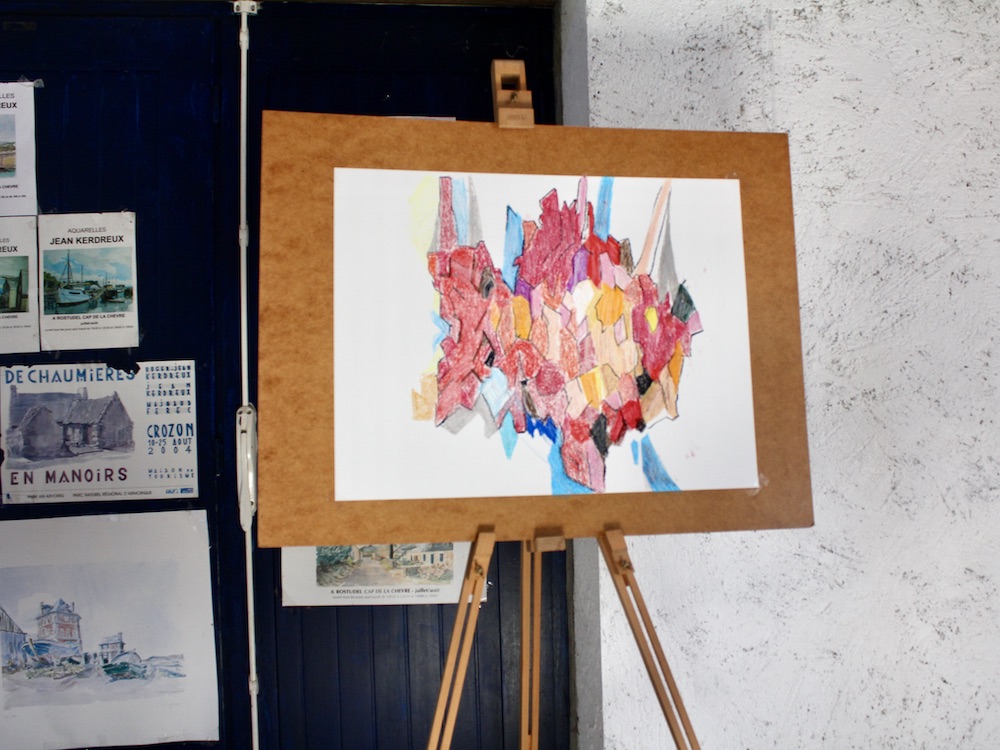}
    \caption{Creative exploitation of perception map. We first ask an artist to chose a stimulus image. We later project on his canvas the same image distorted according to his visual attention (left) or the algorithm. The artist then freely plays with this projection to design a painting (right).}
    \label{fig:creation}
\end{figure}

\paragraph{A creative tool.}
Our renderings are visuals of perceptions, stimuli for thoughts.
Our diptychs could hence inspire artists to represent reality through a new lens, in a similar fashion to \textit{computational catalysts} presented in \cite{kerdreux2020interactive}.
We explore some possibilities in~\cref{fig:creation}.

\paragraph{Ethical Considerations.}
A.I. research is notably inspired by heuristics of cognitive mechanisms that algorithms seek to reproduce.
It has naturally fueled many science-fiction scenarios and myths.
Those narratives are not innocuous as it can correlate with political decisions.
As such, accessible pedagogical content is highly profitable.
Here, we provide illustrations, easily understandable by the many, of some differences between humans and current algorithms. 
This is also helpful in understanding the limits of these algorithms so as to refine them.

\bibliographystyle{plain}
\bibliography{biblio.bib}

\clearpage
\appendix

\section{Additional Images}
\label{app:additional_images}

\begin{figure}[h]
    \centering
    \includegraphics[width=.32\textwidth]{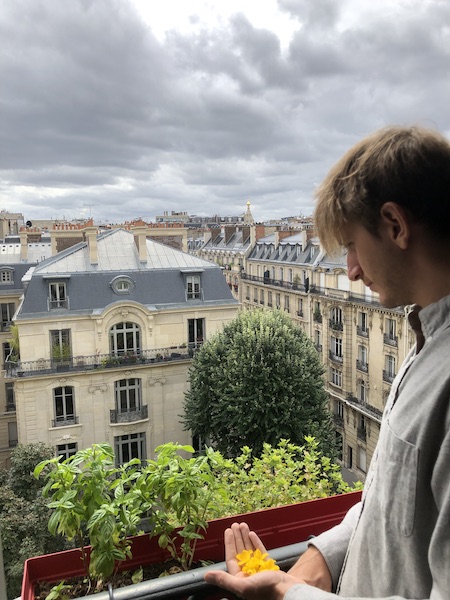} 
    \includegraphics[width=.32\textwidth]{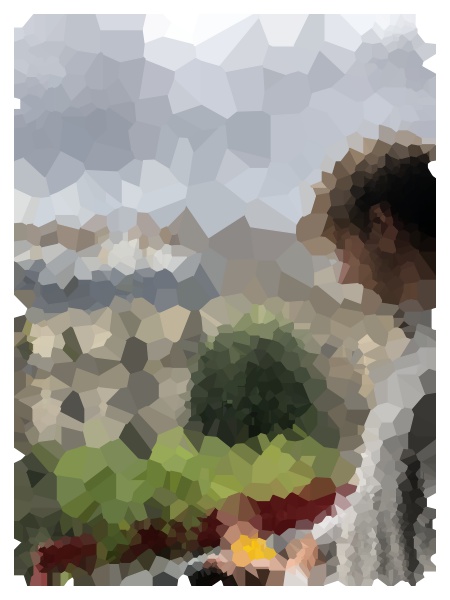} 
    \includegraphics[width=.32\textwidth]{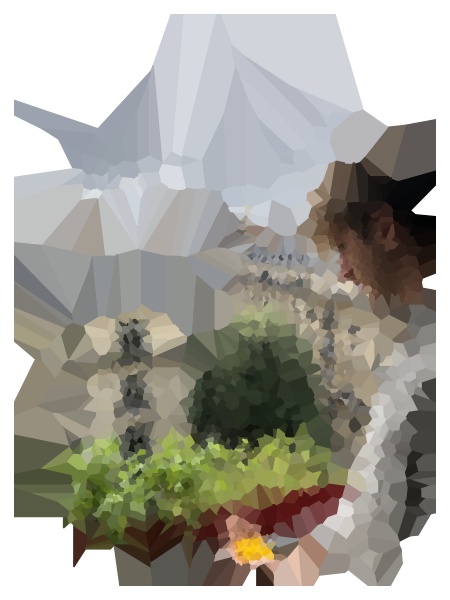}
    
    \includegraphics[width=.32\textwidth]{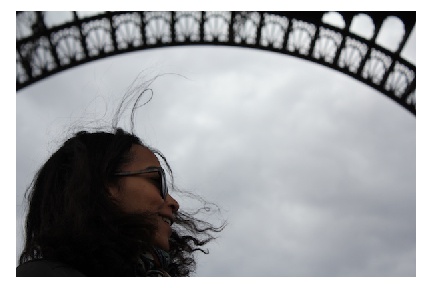} 
    \includegraphics[width=.32\textwidth]{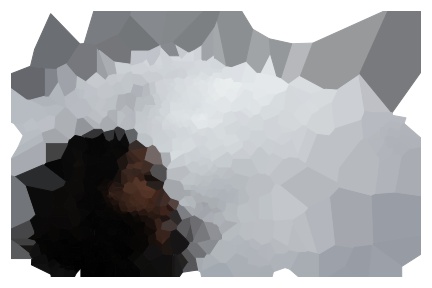}
    \includegraphics[width=.32\textwidth]{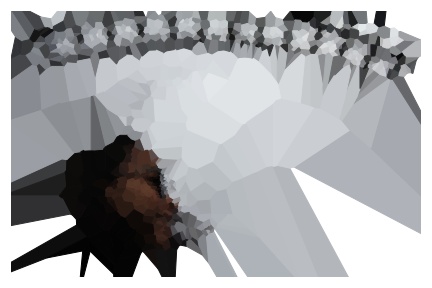}

    \includegraphics[width=.32\textwidth]{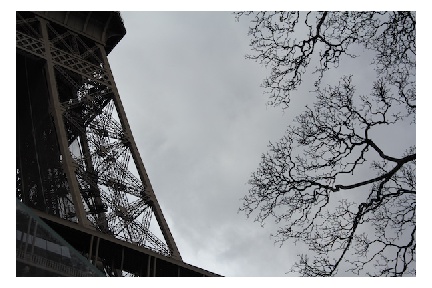} 
    \includegraphics[width=.32\textwidth]{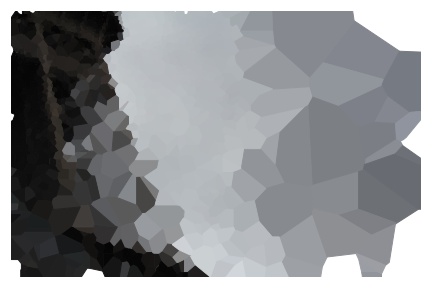} 
    \includegraphics[width=.32\textwidth]{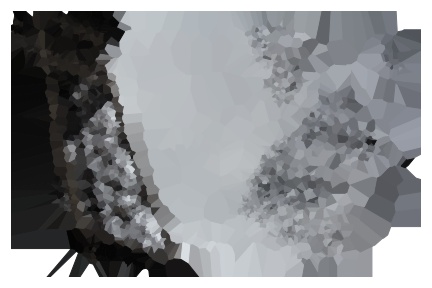}
    
    \caption{Diptychs similar to~\cref{fig:triptychs}. (Top) The first one illustrates the differences in perception between task-oriented AI and the human mind. First of all, human attention, like many animals, is drawn to faces, and most specifically, to the eyes. The human subject also focuses on specific windows below the vanishing point, probably due to contrasts, building symmetry, and picture composition.
    Moreover, the neural network misses a lot of mid-level semantics that structure the image, such as perspective lines on the right, or cucumber blossoms in the foreground. 
    Finally, the human subject seems to examine the basil on the bottom left, reminding us of our contemplative capabilities.
    (Middle) The second diptych also shows human attention given to faces and eyes, even when covered by sunglasses. Note also how our neural network neglects the arch details.
    (Bottom) In the last picture, most of the interest resides in texture differences between the Eiffel tower on the left and trees without foliage on the right, which our neural network, designed to discriminate between \emph{ImageNet} categories, is unfit to grasp.}
    \label{fig:additional}
\end{figure}

\begin{figure}
    \includegraphics[width=.32\textwidth]{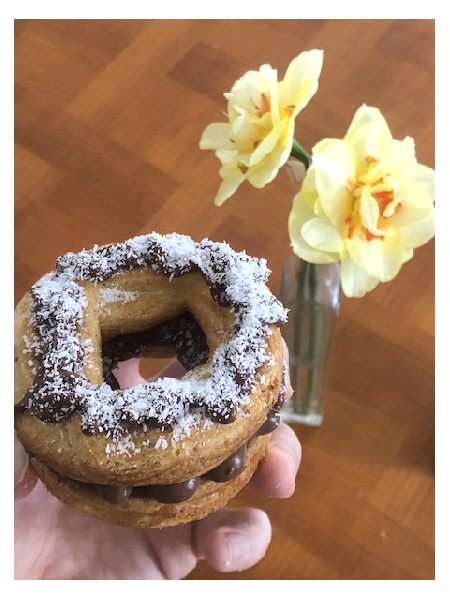} 
    \includegraphics[width=.32\textwidth]{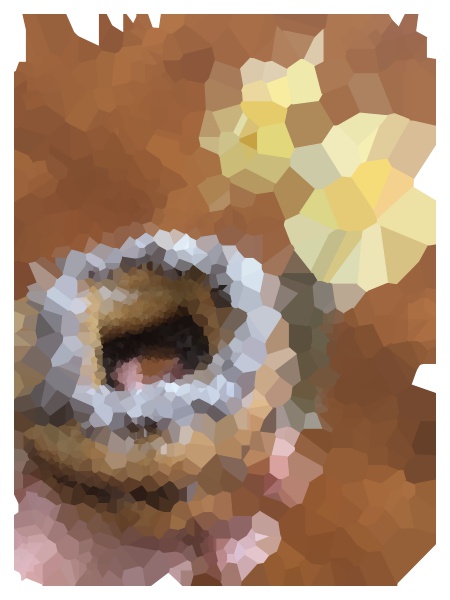} 
    \includegraphics[width=.32\textwidth]{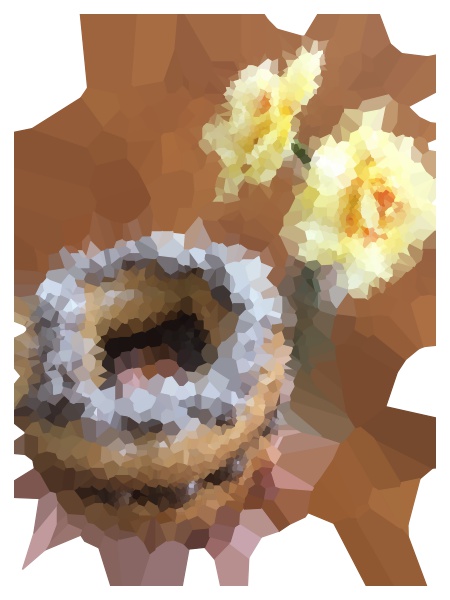} 
    
    \includegraphics[width=.32\textwidth]{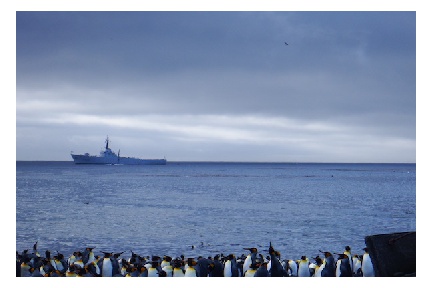} 
    \includegraphics[width=.32\textwidth]{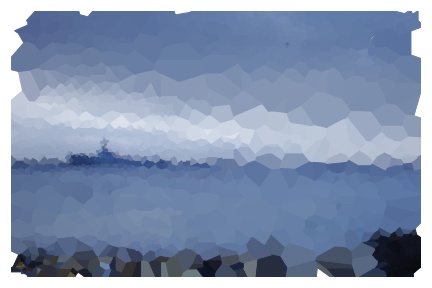} 
    \includegraphics[width=.32\textwidth]{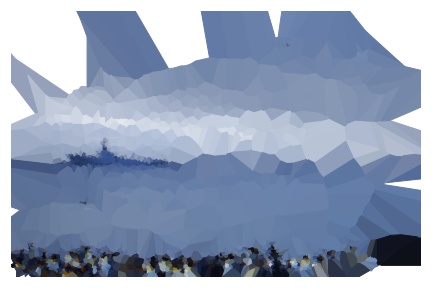}
    
    \includegraphics[width=.32\textwidth]{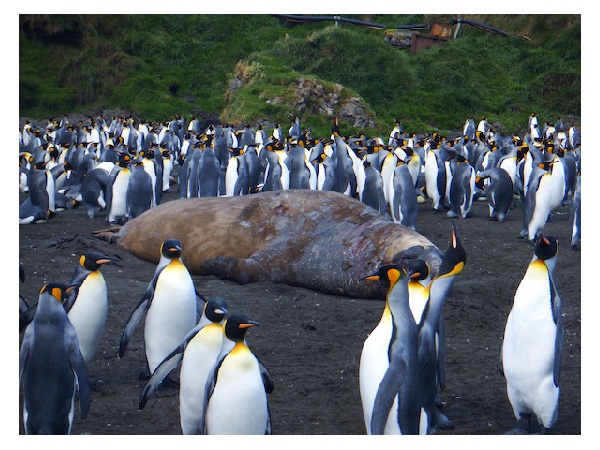} 
    \includegraphics[width=.32\textwidth]{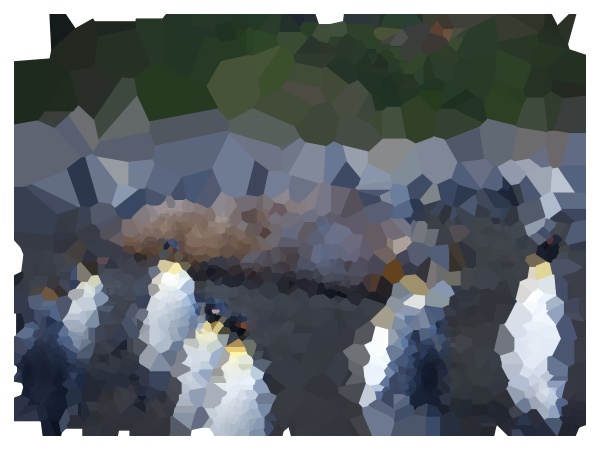} 
    \includegraphics[width=.32\textwidth]{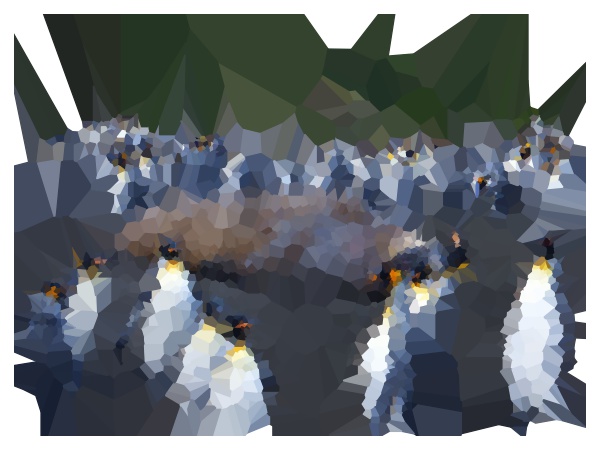}
    
    \caption{This second series of additional diptychs probably better illustrates limitations in current definitions of neural network saliency rather than differences with humans.
    For classification networks, saliency is computed as the gradient of the most probable class with respect to the input. This gradient will be important on the edge of the recognized object, such as the pastry on the first diptych. Yet, due to the non-linearity of the max, this gradient will not be important on the edge of the second most probable object. Note that this could be avoided by taking the sum of the output probably vector rather than the max before computing the gradient with respect to the input.
    Another interesting fact can be inferred from the boat picture. The neural network visual shows that the saliency map is close to being uniform everywhere on this picture, which means that the classifier's output is highly dependent on changes in any regions of the picture, which reminds us of intriguing properties of neural networks that were brought to light by adversarial examples \cite{42503}.}
    \label{fig:two_objects}
\end{figure}

\clearpage

\section{Additional Considerations}
\label{app:choices}

\paragraph{Visual aesthetics.}
While designing our visual diptychs, we have made a few choices that we will discuss here briefly. 
First of all, if we went for a mosaic rendering, one could enhance saliency with other methods than our tiling one. For example, one could use impressionist or pointillist style with different brush sizes, allowing to relay more details on salient parts of the image.
Secondarily, our procedure voluntarily leads to tiles crossing over the edges of the original image, creating fuzzier images, which perturbs our usual object-oriented perception.
Finally, frontier regions are left blank, underlying that our perceptions discard a vast amount of our sensual stimuli.

\begin{figure}[hb]
    \centering
    
    \includegraphics[width=.3\textwidth]{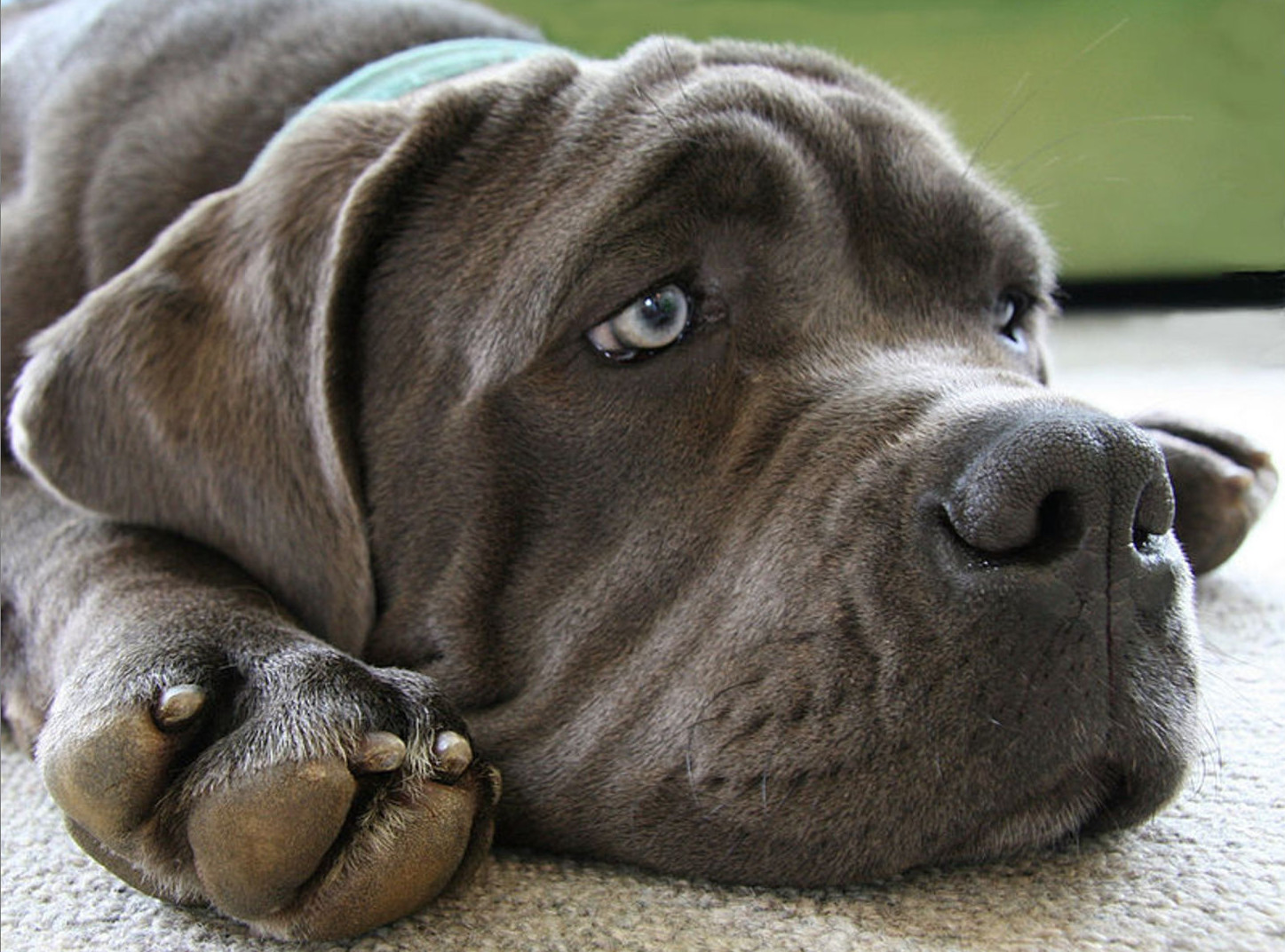} 
    \includegraphics[width=.3\textwidth]{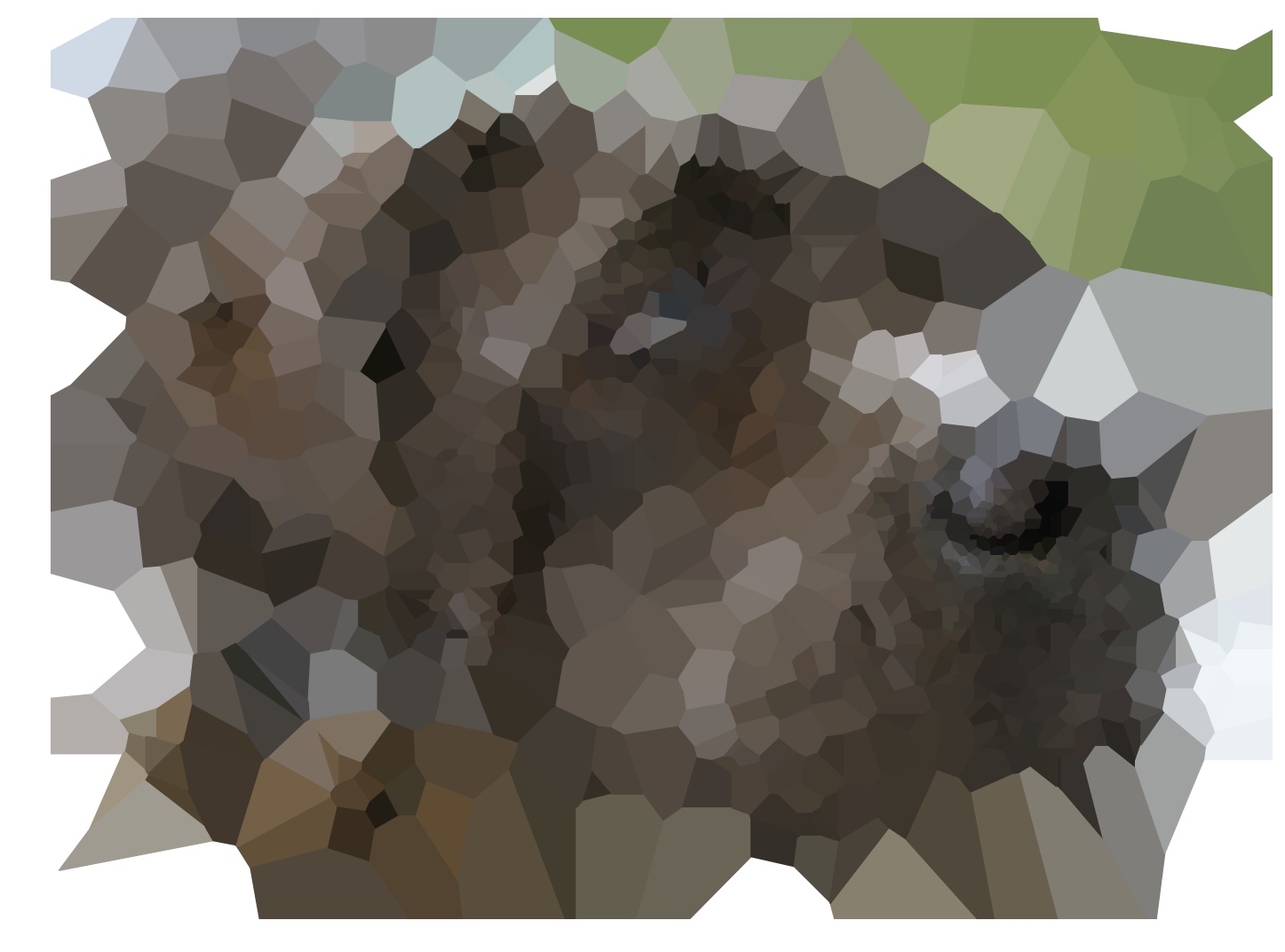}
    \includegraphics[width=.3\textwidth]{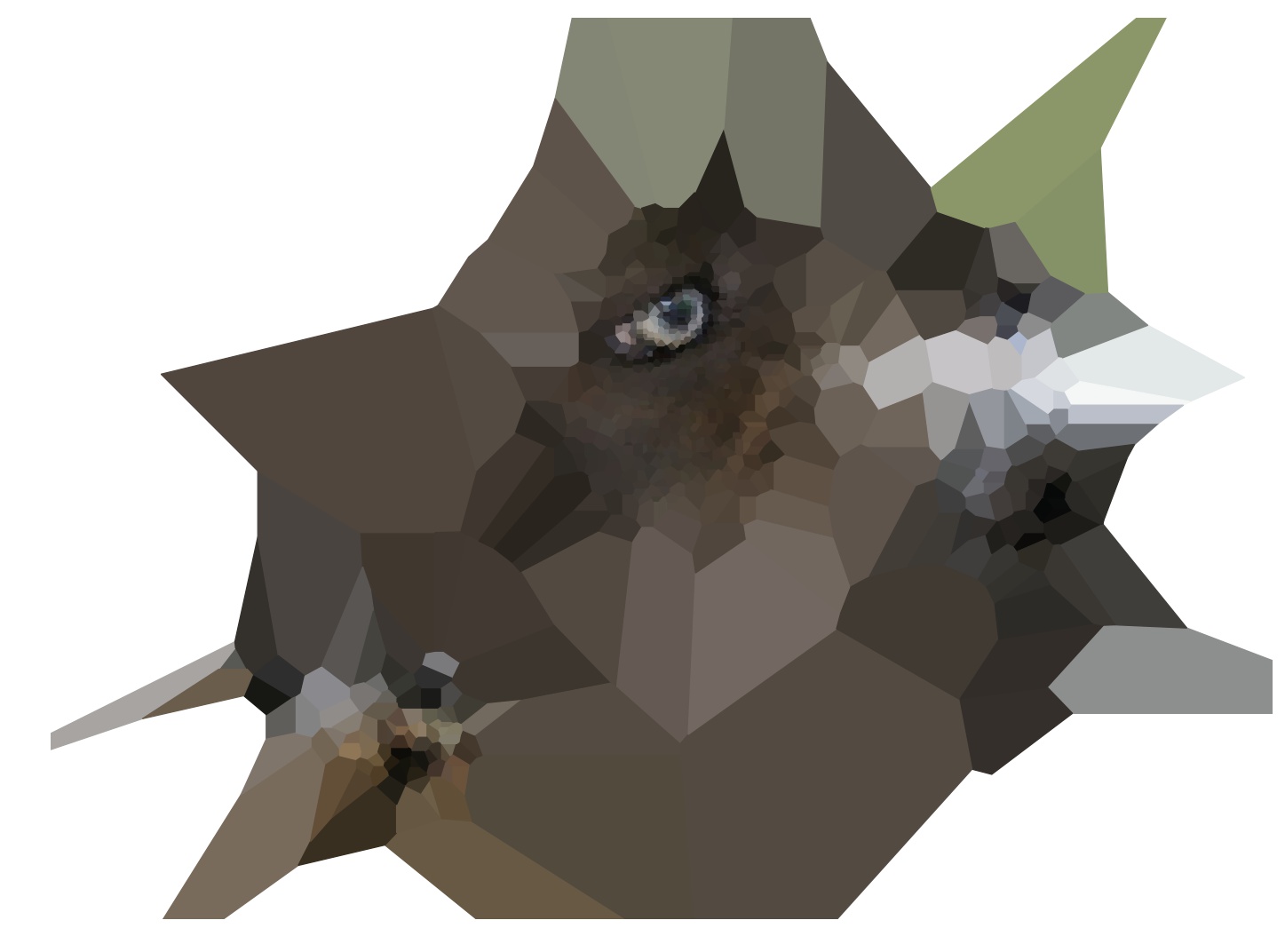}
    
    \vspace{2mm}

    \includegraphics[width=.3\textwidth]{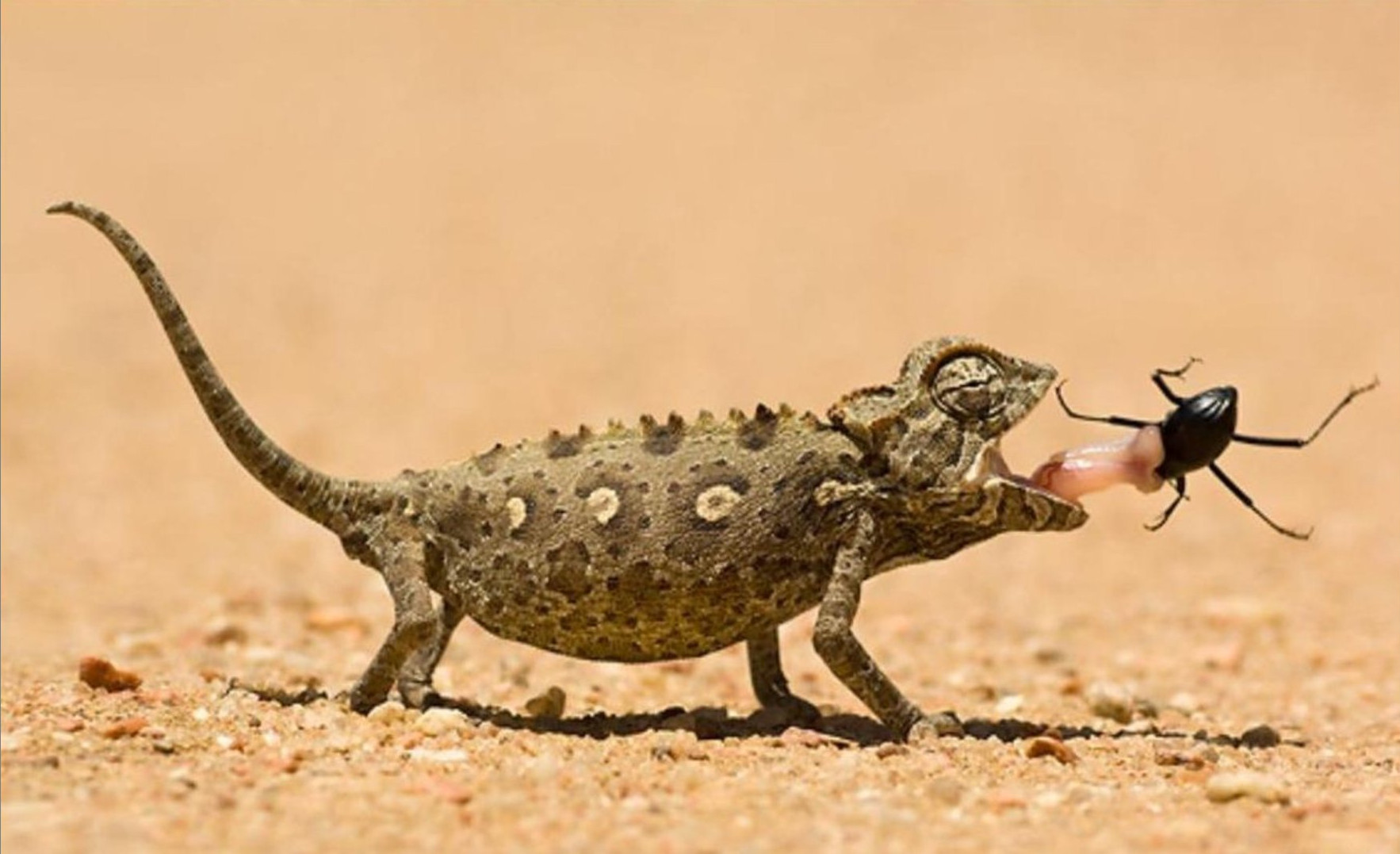} 
    \includegraphics[width=.3\textwidth]{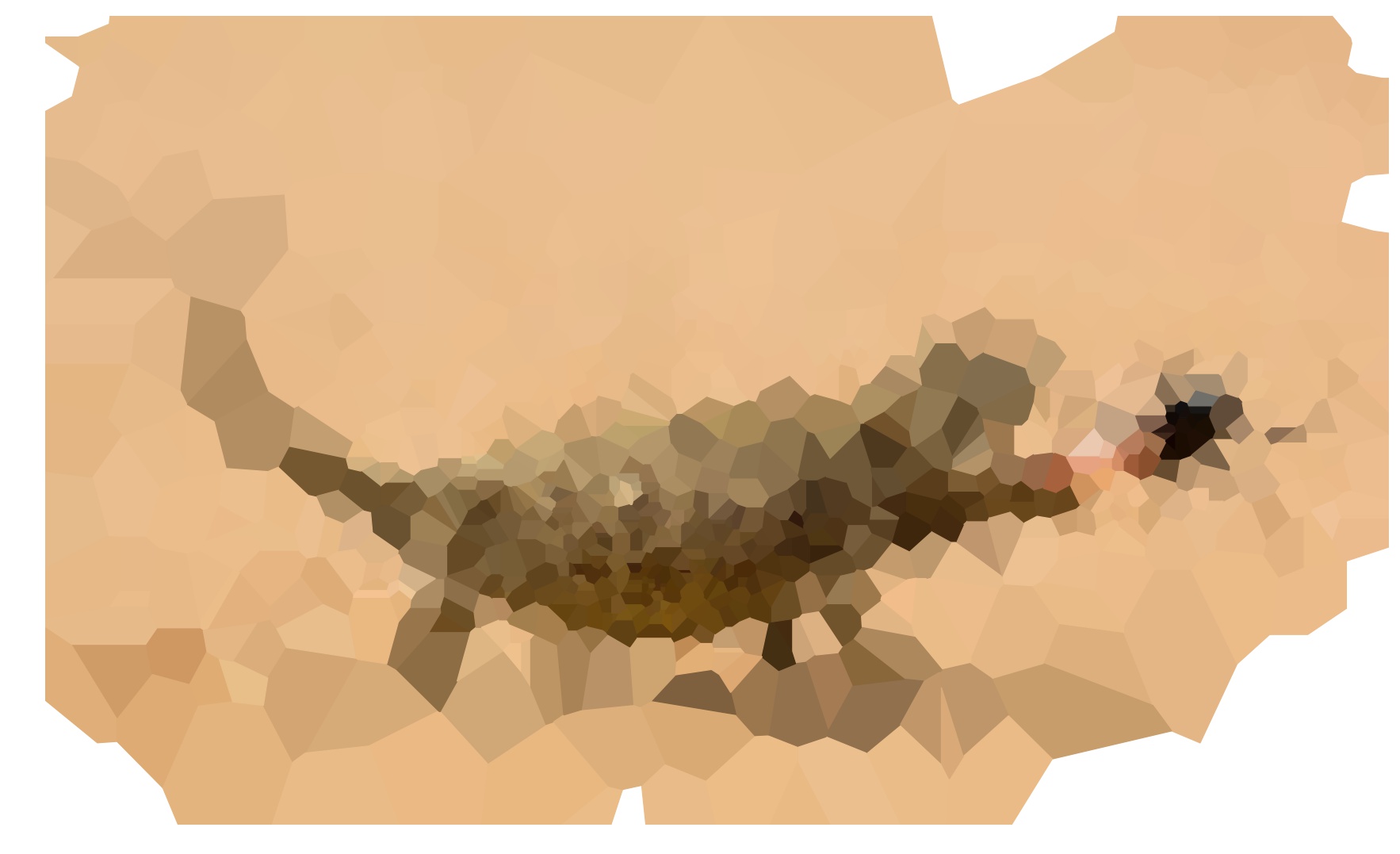}
    \includegraphics[width=.3\textwidth]{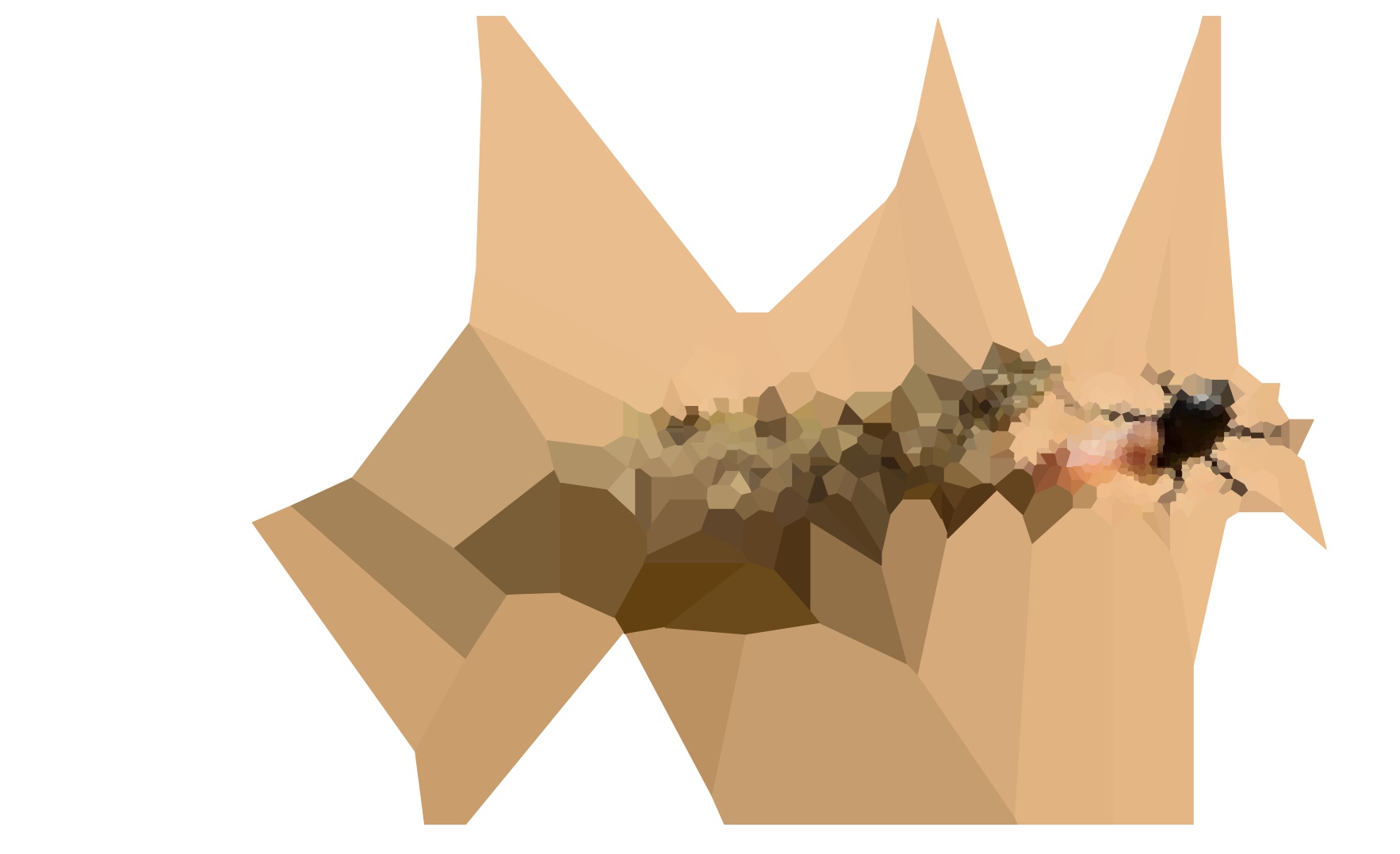}
    
    \vspace{2mm}
    
    \includegraphics[width=.3\textwidth]{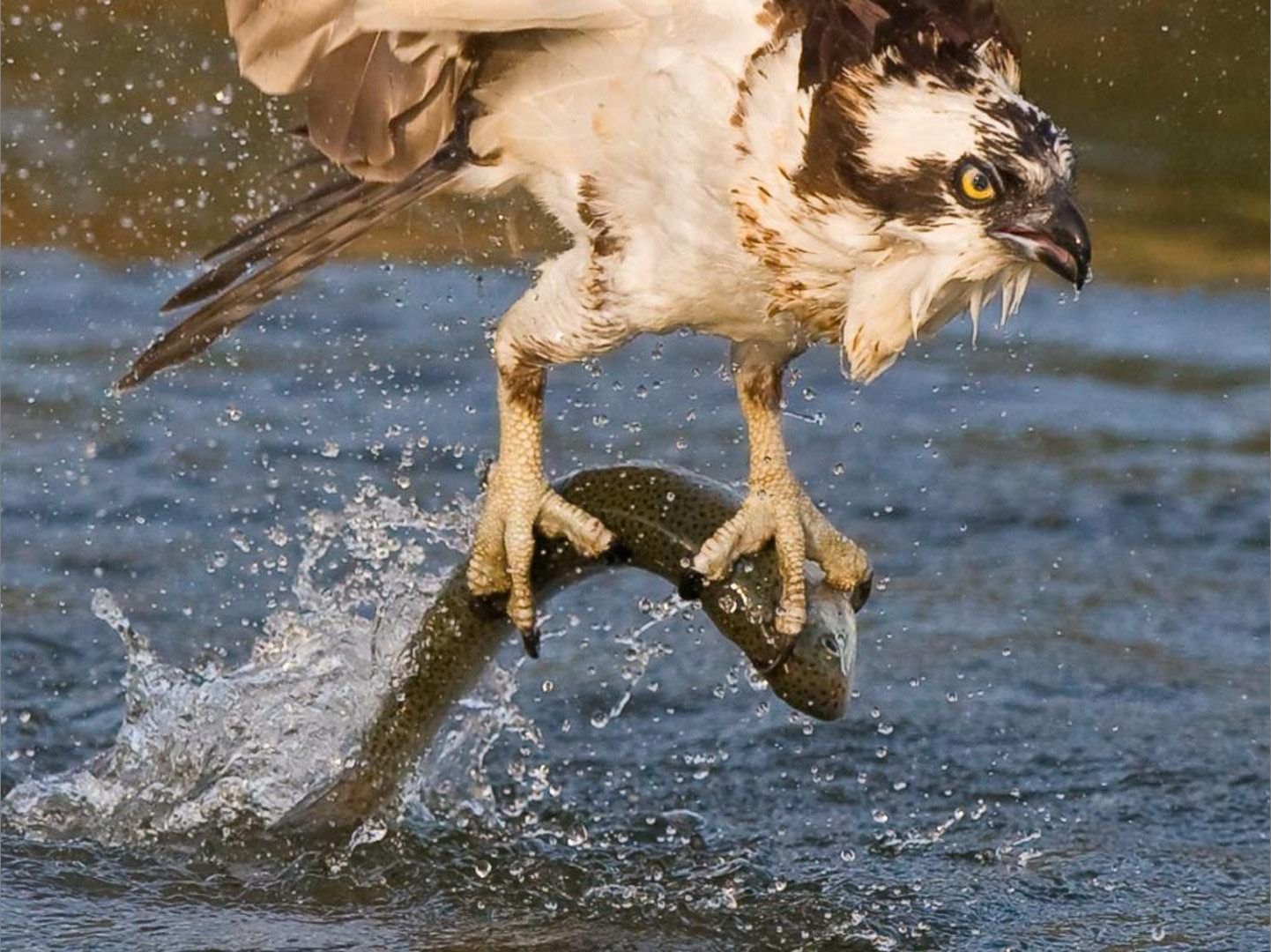} 
    \includegraphics[width=.3\textwidth]{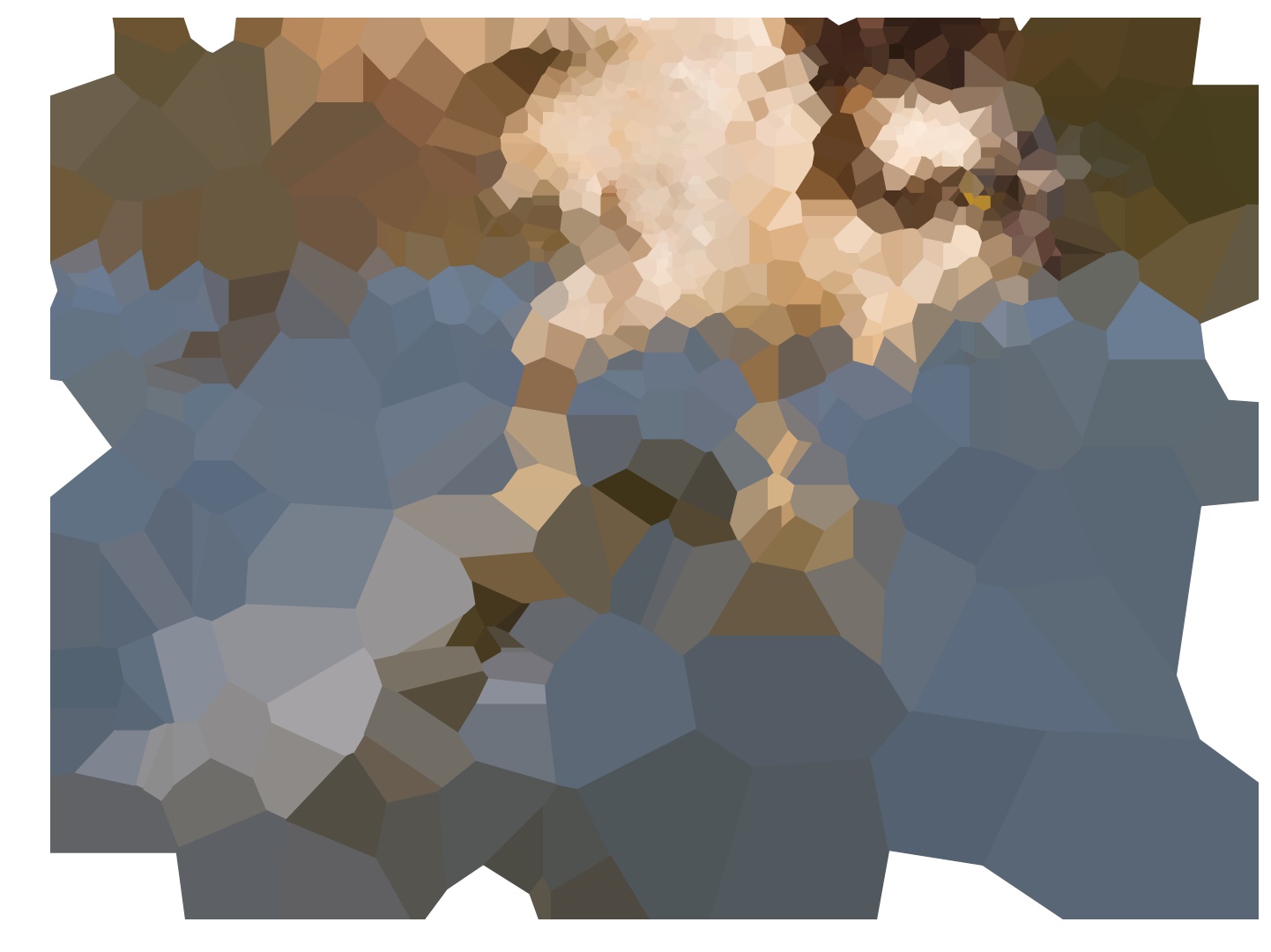}
    \includegraphics[width=.3\textwidth]{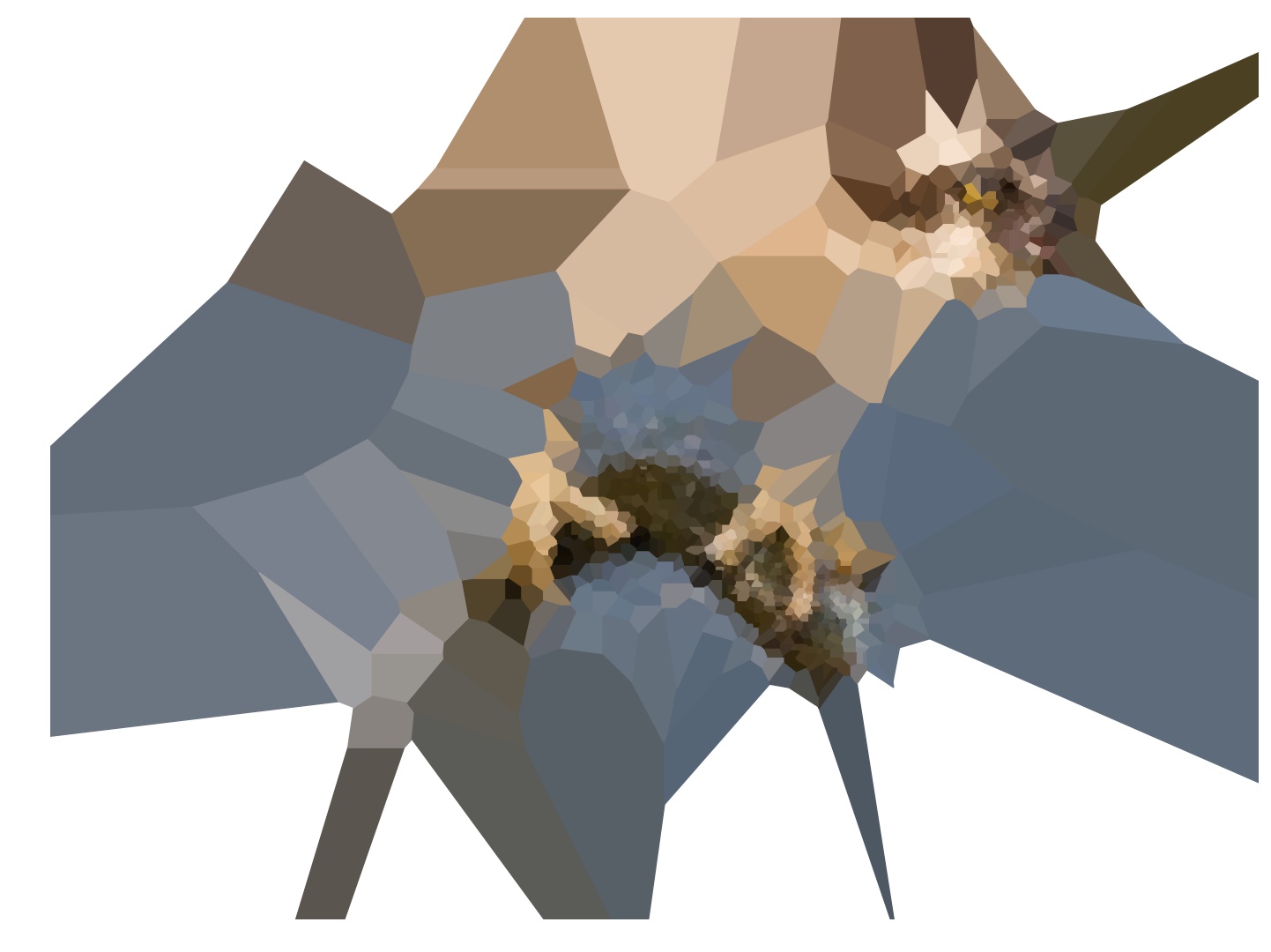}
    
    \caption{Diptychs similar to~\cref{fig:triptychs} when human saliency maps are retrieved from \emph{CAT2000} database.
    On those images, one can see how the human focus is drawn on faces with particular attention to eyes while the neural network perceives silhouettes uniformly.
    Moreover, for both the chameleon and the eagle images, our classifying neural network does not pay attention to the hunting action.}
    \label{fig:cat2000}
\end{figure}

\paragraph{Measuring human saliency.}
Defining and measuring saliency is object to debate. In this section, we aim to illustrate that the eye tracking system we have designed to retrieve fixation maps for human, does not lead to results that are inconsistent with existing reviews of human saliency found in the literature. 
There exists many different databases of human attention on images, each emphasizing specific type of images, such as faces \cite{cerf2009faces}, semantically affective scenes \cite{ramanathan2010eye}, crowds \cite{jiang2014saliency} and many others \cite{judd2011fixations,xu2014predicting}. 
In~\cref{fig:cat2000}, we abandon our human eye tracking system, using fixation maps from the \emph{CAT2000} database \cite{borji2015cat2000}. It illustrates the robustness of our visuals to the measure of human saliency.

\end{document}